\documentclass[aps, twocolumn,floats, showpacs]{revtex4}
\usepackage{amsmath,amssymb,graphicx}
\usepackage{epsfig}
\usepackage{graphicx}
\usepackage{enumerate}

\newcommand{\beq}{\begin{equation}}
\newcommand{\eeq}{\end{equation}}

\newcommand{\bea}{\begin{eqnarray}}
\newcommand{\eea}{\end{eqnarray}}

\begin{document}
\title{ Thermoelectric effect in molecular junctions: A tool for revealing of transport mechanisms. }
\author{Dvira Segal}
\affiliation{Department of Chemical Physics, Weizmann
Institute of Science, 76100 Rehovot, Israel.}

\date{\today}
\begin{abstract}

We investigate the thermopower of a metal-molecule-metal junction
taking into account thermal effects on the junction.
Based on analytical expressions and numerical simulations
we show that the thermoelectric potential reveals valuable information on the
mechanisms controlling the electron transfer process,
  including coherent transmission and thermalized hopping.
We also show that at high temperatures the position of the Fermi energy
relative to the molecular states
can be easily deduced from the thermoelectric potential.
Standard current-voltage measurements are insensitive to this information.

\end{abstract}

\pacs{
73.40.-c,
73.63.-b,
85.65.+h,
73.50.Lw
}

\maketitle

\section{ Introduction}

Understanding of charge transfer processes through single
molecules is at the forefront of molecular electronics
\cite{MoleR}. The electrical conductance of a single molecule
coupled to metal electrodes has been recently measured by
different techniques \cite{IVexp1, IVexp2}. In these
experiments the current-voltage (I-V)  characteristic of the
device is measured, and information on the transport mechanism,
the molecule-metal coupling strength,  and the role played by nuclear motion
in the conduction process are deduced.

Experimental data \cite{Ratner, DNA, Selzer} and theoretical
studies \cite{Emberly, Ness, Segal, Todorov, Petrov} suggest that
two mechanisms are involved in electron transfer processes through
molecular bridges: Super-exchange mechanism and thermal induced
hopping. The super-exchange mechanism is a coherent (tunneling)
and short distance charge-transfer process. It dominates transport
when the  molecular levels relevant for transport- Fermi level
offset is greater than the thermal energy. In the opposite limit
electron transmission occurs through sequential hopping along the
bridge. This is a multistep process that allows long distance
charge transfer. In a typical I-V experiment transport mechanisms
are analyzed based on Arrhenius plot, revealing a characteristic
transition from temperature ($T$) independent behavior
 at low $T$ to a strong dependence at high $T$ \cite{Selzer}.

It was recently suggested that thermoelectric measurements could
provide additional insight into electron transport through {\it
single} molecules, and contain information on the electronic and
vibrational excitation spectrum of the molecule \cite{Datta,
Zheng, Oreg}.

In a thermoelectric experiment the electric current induced by a
finite temperature difference is investigated. Alternatively, the
potential under the condition of zero current is measured. The
thermoelectric power of microstructures such as quantum dots
\cite{QD}, single electron transistors, \cite{SET} and mesoscopic
nanotubes \cite{CNT1, CNT2} has been of interest since it yields
directly the sign of the dominant charge carriers and the
intrinsic conduction properties. In addition, it is sensitive to
the electronic structure at the Fermi level.
In two recent studies the thermoelectric voltage over a conjugated
molecular conductor \cite{Datta} and an atomic chain \cite{Zheng}
was calculated. It yields valuable information- the location of
the Fermi energy relative to the molecular levels. Inelastic
interactions on the bridge were neglected in both cases. In a
different work Koch {\it et al.} \cite{Oreg} investigated the
thermopower of a single molecule when both sequential tunneling
(lowest order tunneling process) and cotunneling (second order
tunneling processes) take place. Vibrational features of the
molecule were taken into account, but direct thermal activation of
electrons onto the bridge and the possibility of diffusional
transport were not included. In addition, this study was limited
to the shortest molecular unit made of a single electronic state.

In this paper we extend these ideas and analyze the thermoelectric
voltage of a molecular junction of bridge length $N>1$ while
taking into account both coherent and thermal interactions in a
unified theory. Our model includes relaxation mechanisms on the
bridge, arising from the interaction of the electronic system with
other bridge or environmental degrees of freedom. The effect of
these interactions is to open up a thermal channel for conduction.
Hence in our model electrons can be transmitted through the bridge
either coherently (tunneling), or by sequential hopping. We focus
here on the nonresonant regime, where bridge energies are far
above the chemical potentials of the metals. In this limit the
actual population of electrons on the bridge is very small, and
effects due to charging of the bridge (coulomb blockade) are
negligible.

We show through simple approximate expressions and with numerical
examples that the functional behavior of the thermoelectric
voltage is intrinsically different for the different modes of
transfer, and that it clearly reflects the turnover between the
two mechanisms. The thermoelectric phenomena can therefore serve
as a significant tool for analyzing transport mechanisms
complementing traditional techniques. In addition, we find that
when thermal interactions dominate the transport, the molecular
levels -Fermi energy gap for transmission can be easily deduced
from the thermopower value. This last property is crucial for
interpreting I-V results in transport experiments.


\section{ Model}

\begin{figure}[htbp]
\vspace{-20mm}
\hspace{0mm}
 {\hbox{\epsfxsize=110mm \epsffile{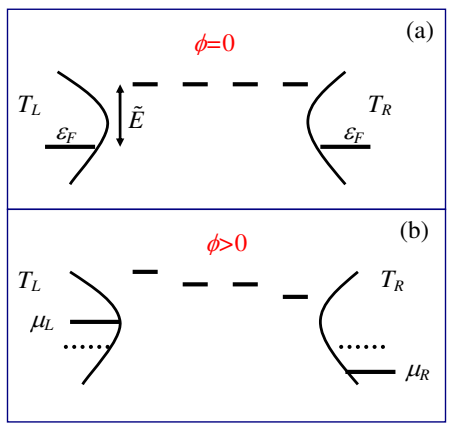}}}
\vspace*{-80mm}
 \caption{(Color online) Scheme of the model system: A molecule of $N$ sites connecting two metal leads.
(a) No applied bias. (b) $e\phi>0$.}
 \label{figmodel}
\end{figure}

The model system consists of a metal-molecule-metal junction under
an electrical bias and a thermal gradient. The molecule is
described by a tight binding model with $N$ sites with one state
localized at each site. The first and last states are coupled to
the left ($L$) and right ($R$) metal leads respectively. At
equilibrium, $\epsilon_F$ specifies the Fermi energy of the two
 metals, taken as the zero of energy.
  Under a potential bias $\phi$ the leads are characterized by electrochemical potentials
$\mu_L$ and $\mu_R$ for the $L$ and $R$ sides. In addition, the
two metals are kept at different temperatures $T_L$ and $T_R$. For
a schematic representation see Fig. \ref{figmodel}.

We investigate the thermoelectric effect in this system using a variation
of the model developed before to analyze thermal effects in
electron transmission through molecular bridges \cite{Segal, Segalh}.
 There, the molecular system and the metals were
 in contact with a {\it single} thermal reservoir held at temperature $T$.
Here, in contrast, we specify only the temperatures of the charge
reservoirs (metals), and from this boundary condition the
molecular temperature should be deduced.
We explain below how to implement this modification into the
formalism of Ref. \cite{Segal}.

The system Hamiltonian includes five terms
\beq H= H_M + H_K + H_{MK} +  H_{B} + H_{MB}. \label{eq:Ham}
\eeq
Here $H_M$ denotes the isolated molecule of length $N$  with one
electronic state at each site
\beq
 H_M=\sum_{j=1}^{N}E_j|j\rangle\langle j| +
 V\sum_{j=2}^{N}\left( |j\rangle \langle j-1| +  |j-1\rangle \langle j|\right).
\eeq
The bridge energies $E_j$ are taken to be equal at all sites at
zero bias, $E_j=\tilde {E}$. $V$ is the nearest neighbors
electronic coupling.
The left and right charge carriers reservoirs (metals)
are described by their set of electronic states
\beq H_K=\sum_{r\in R}\epsilon_r |r\rangle \langle r | +
\sum_{l\in L}\epsilon_l|l\rangle \langle l |. \eeq
The leads are held at constant temperatures $T_L$ (left) and $T_R$
(right). In what follows we use the notations $T_a=(T_L+T_R)/2$
and $\Delta T=T_L-T_R$.
The molecule-metals interaction term is
\beq H_{MK}=\sum_{l \in L}V_{l,1}|l\rangle\langle1| + \sum_{r \in
R}V_{r,N}|r\rangle\langle N | +c.c. \,\ , \eeq
yielding the relaxation rate $\Gamma_L(\epsilon)=2\pi\sum_{l \in
L} |V_{1,l}|^2\delta(\epsilon-\epsilon_l) $, and an analogous
expression for $\Gamma_R$ at the $R$ side.

The molecule is in contact with thermal degrees of freedom, both
internal (vibrations), and those related to the motion relative to
the leads, all denoted as a "thermal bath" and included in $H_B$.
The last term in the Hamiltonian (\ref{eq:Ham}) describes the
coupling of this thermal bath to the molecule. Specifically we use
the following model for this interaction
\beq H_{MB}=\sum_{j=1}^{N}F_{j}|j\rangle \langle j|,
\label{eq:HMB}
 \eeq
where $F_j$ are bath operators and $|j\rangle$ are molecular
states, $j=1...N$.
The form of Eq.~(\ref{eq:HMB}) implies that the thermal bath induces
energy dephasings on the local bridge sites.
The bath operators are characterized by their time
correlation function
\beq \int_{-\infty}^{\infty} e^{i\omega t}\langle F_j (t)
F_{j'}(0)\rangle=e^{\beta \omega} \int_{-\infty}^{\infty}
e^{i\omega t}\langle F_{j'} (0) F_{j}(t)\rangle, \label{eq:det}
\eeq
where $\beta=1/k_BT$, $T$ is the local thermal bath temperature and $k_B$
is the Boltzmann constant.
 Assuming that thermal interactions on
different molecular sites are not correlated, and going into the
markovian limit, the correlation function becomes
\beq \langle F_j (t) F_{j'}(t')\rangle
=\gamma\delta_{j,j'}\delta(t-t').
\eeq
Here $\gamma$ is the dephasing rate reflecting the
strength of the system-bath interaction: When $\gamma=0$, the
electronic system and the thermal degrees of freedom are
decoupled, and electrons move coherently along the wire. In
contrast, strong dephasing rates imply the dominance of the
incoherent transmission mode \cite{Segal}. It should be emphasized
that the bridge dephasing rate $\gamma$ and the electronic
couplings between the sites $V$ are in general temperature
dependent. Since in the nonresonant regime thermal activation is
the dominant temperature dependent factor, we simplify the
discussion and ignore these corrections. The thermal bath
temperature therefore enters the formalism only through the
detailed balance condition, Eq. (\ref{eq:det}).

The transport behavior of the system depends, among other factors,
on the way the potential bias falls along the molecular bridge. In
what follows we use the following model for the electrostatic
potential profile
%
\bea \mu_L&=&\epsilon_F+e\phi/2; \,\ \mu_R=\epsilon_F-e\phi/2;
\nonumber\\
E_1&=& \tilde{E}+e\phi/4; \,\ E_N=\tilde{E}-e\phi/4; \,\
\nonumber\\
E_{j}&=&\tilde{E}, \,\ j=2..N-1. \label{eq:mu}
 \eea
The qualitative aspects of the results presented below are not
affected by this particular choice, since the relation $\Delta E
\equiv (\tilde E-\epsilon_F)>e\phi$ is retained in this work.

In the weak molecule-thermal bath coupling limit a computational
scheme for evaluating the energy dependent transmission
coefficient through a metal-molecule-metal junction for the
$T_L=T_R$ case was developed in Ref. \cite{Segal}. The method is
based on the generalized quantum master equations extended to
steady state situations. Here we further extend this framework and
study thermoelectric effects in molecular junctions.

\section{Thermopower}

The current through the junction is calculated by generalizing the
Landauer formula \cite{Landauer} to situations involving inelastic
interactions on the bridge \cite{bonca, Nitzan}

\bea
 I&=&I_{{L\rightarrow R}} -I_{R\rightarrow L},
\nonumber\\
I_{L\rightarrow R}&=&\frac{e}{\pi \hbar} \int d\epsilon_0 \int
d \epsilon_f \mathcal T_{L \rightarrow R}(\epsilon_0,
\epsilon_f)f_L(\epsilon_0) \left( 1-f_R(\epsilon_f)\right),
\nonumber\\
I_{R\rightarrow L}&=&\frac{e}{\pi \hbar} \int d\epsilon_0 \int
d \epsilon_f \mathcal T_{R \rightarrow L}(\epsilon_0,
\epsilon_f)f_R(\epsilon_0) \left( 1-f_L(\epsilon_f)\right).
\nonumber\\
\label{eq:curr}
 \eea
Here $\mathcal T_{L \rightarrow R}(\epsilon_0, \epsilon_f)$
denotes the transmission probability for an electron incoming from
the left lead at the energy $\epsilon_0$ to be emitted at the
opposite side at $\epsilon_f$. The energy difference
$\epsilon_f-\epsilon_0$ is exchanged with the thermal environment.
This coefficient depends on the molecular parameters, the applied
potential, the metal-molecule interaction, temperature and
dephasing. It is not necessarily the same for the different
directions. The current (defined positive when flowing left to
right) is evaluated by weighting the transmission probability by
the appropriate combination of the Fermi distribution functions at
the metals $f_K(\epsilon)=\left(e^{\beta_K(\epsilon-\mu_K)}+1
\right)^{-1}$ where $\beta_K\equiv (k_BT_K)^{-1}$ is the inverse
temperature ($K=L,R$). Unless specified, the integrals are all
taken as $\int_{-\infty}^{\infty} $.

The generalized Landauer equation does not take into account the
effect of the contact population on the inelastic processes.
Including such effects can be implemented by replacing the Fermi
occupation factors by {\it nonequilibrium} electron distribution
functions \cite{Emberly}. Yet it can be shown that Eq.
(\ref{eq:curr}) well approximates the current when the
transmission
 through the junction is significantly small ($<<1$)
\cite{antonyuk}. Since we focus here on the out of resonance,
small bias, weak electron-phonon interaction and weak
metal-molecule coupling situation, transmission probabilities are
always small and Eq.~(\ref{eq:curr}) can be utilized.
The inelastic Landauer formula can be also derived using the
systematic nonequilibrium Green function approach. Recent
calculations of inelastic tunneling currents yield an expression
similar to (\ref{eq:curr}) assuming weak coupling to the leads,
weak electron-phonon coupling, and small bias \cite{MishaJPC}. For
recent discussions on the issue see Refs. \cite{Oreg,domcke,
antonyuk}.



%
\begin{figure}[htbp]
\vspace{0mm} \hspace{0mm}
 {\hbox{\epsfxsize=70mm \epsffile{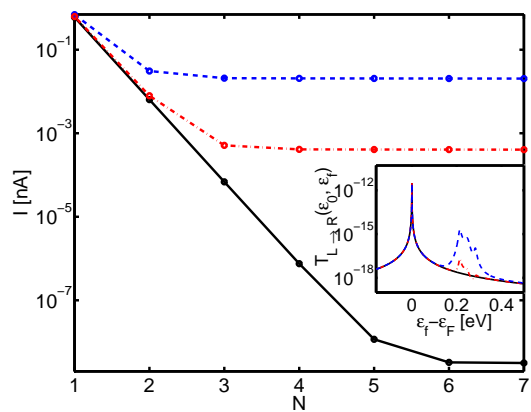}}}
 \caption{ (Color online) Electron current vs. wire length $N$ for different temperatures:
 $T$=300 K (dashed), $T$=200 K (dashed-dotted) and $T$=100 K (full).
The other parameters are $\Delta E$=250 meV, $V$=25 meV, $e
\phi$=2.5 meV, $\Gamma_L=\Gamma_R$=15 meV, $\gamma$=1 meV. The
inset presents the respective $L$ to $R$ transmission probability
$\mathcal T_{L \rightarrow R}$ for $N$=3 and
$\epsilon_0=\epsilon_F$. $T$=300 K ,200 K, 100 K top to bottom. }
 \label{Fig0}
\end{figure}

We calculate next the current flowing through a biased $N$ level
molecular junction. We first consider the simple situation of
$T_L=T_R=T$, i.e. we do  not apply an external temperature
gradient. Since this system was analyzed extensively in previous
works, Refs. \cite{Segal, Segalh}, we present here only the main
results. Fig.~\ref{Fig0} demonstrates the distance dependence of the
current at different temperatures. The following set of parameters
is used: $\Delta E=250$ meV, $V=25$ meV, $e\phi=2.5$ meV and
$\gamma=1$ meV (Typical dephasing times in condensed phases are of
order of 1 ps).
The inset depicts the corresponding energy resolved transmission
probability for an electron incoming at $\epsilon_0=\epsilon_F $
for the $N=3$ case.

The following observations can be made: (i) The transmitted flux
consists in general two main components: elastic tunneling at
$\epsilon_f$= $\epsilon_0$ and activated flux in the energy range
of the bridge states $\sim \Delta E$. These two components can be
clearly distinguished when bridge energies are high enough,
 $\Delta E\gg k_BT$ and the coherent interaction
 $V$ is much smaller than $\Delta E$.
(ii) The coherent component is most important at low temperatures,
 large energy gaps and short chains. The incoherent contribution dominates
 in the opposite limits.
(iii) The electric current manifests a clear transition from an
exponential decay typical to tunneling for short chains, to a weak
(algebraic) distance dependence for longer chains, characterizing
thermal activation into the bridge. The turnover is shifted into
higher $N$ values when the temperature is lowered.
(iv)  The thermal component  of the transmission is exponentially
enhanced when increasing the temperature. In contrast, the effect
of the dephasing is more subtle. When the dephasing rate is very
small $\gamma < 0.01$ meV (using the parameters of Fig.
\ref{Fig0}), transport is coherent and the current increases with
$\gamma$. For very high dephasing values, $\gamma\sim 100$ meV,
coherent effects are completely destroyed and the current {\it
goes down} like $1/\gamma$. See Ref. \cite{Segal} for details. In
this work we employ the intermediate values of $\gamma\sim 0.1-10$
meV.

Following these observations we can approximate the transmission
coefficient by a generic functional form containing only its main
features: For large energy gaps and for reasonable temperatures
mixed coherent-inelastic contributions can be safely neglected and
the transmission is approximated by two separate terms
\cite{Segal, Segalh}
 \beq \mathcal T= \mathcal T_{tunn}+\mathcal
T_{hopp}.
\label{eq:Trans} \eeq
The first term stands for tunneling
\beq \mathcal T_{tunn}(\epsilon_0, \epsilon_f)=\delta(\epsilon_0-
\epsilon_f)A(\epsilon_0), \label{eq:Ttunn}
 \eeq
while the second contribution reproduces thermalized hopping
through the junction
\beq \mathcal T_{hopp}(\epsilon_0, \epsilon_f)=
B(\epsilon_0,\epsilon_f) e^{-\beta(E_B-\epsilon_0)}.
\label{eq:Thopp} \eeq
$E_B$ is an effective bridge energy. It is related closely to the
gap $\Delta E$ of the molecule in an equilibrium situation, but it
is modified by the molecule-metal interactions and the applied
potential. The coefficients $A$ and $B$ depend on the
molecule-metal coupling terms, the energetics of the bridge, its
length, and on the thermal parameters $T$ and $\gamma$. Here we
assume that $A$ does not depend on the thermal parameters, and
that $B$ is the same when exchanging between the initial and final
energies, $B(\epsilon_0, \epsilon_f)=B(\epsilon_f, \epsilon_0) $
\cite{Segal}. In what follows we utilize Eqs.~(\ref{eq:Ttunn})-
(\ref{eq:Thopp}) for deriving simple expressions for the
thermopower coefficient.


We proceed to the relevant case of an electrically biased junction
under an additional temperature gradient. In our model the
temperatures of the two leads $L$ and $R$ are kept fixed at $T_L$
and $T_R$ respectively, while the temperature of the molecular
degrees of freedom is adjusted to the boundary conditions:
At a steady state situation the
temperature distribution along the molecule is determined, among
other factors, by the metals temperatures, the rate of energy
dissipation on the molecule, and the rate at which energy is
transferred away from the conductor \cite{Segalh, Segalc}. Here we
do not calculate this temperature gradient, but assume that at the
left end of the molecular chain the temperature is nearly $T_L$,
and similarly at the right edge it is close to $T_R$.

We assume next that the transmission $\mathcal T_{K\rightarrow
K'}$ depends on the temperature  $T_K$ ($K$=$L$,$R$), but is
independent of $T_{K'}$. Under this approximation the transmission
coefficient in Eq.~(\ref{eq:curr}) can be calculated using the
procedure of Refs. \cite{Segal, Segalh} without modifications,
simply by employing the different temperatures when evaluating
$\mathcal T _{L\rightarrow R}$ and $\mathcal T_{R \rightarrow L}$.
This approximation relies on the fact that in the $K\rightarrow K'
$ transmission process the temperature at the $K$ end dominates
the transport. This is true considering both transport mechanisms,
tunneling and sequential hopping: Hopping through the bridge is
triggered by thermal activation from the metal, see
Eq.~(\ref{eq:Thopp}). Assuming that this is the rate determining
step, the relevant temperature is therefore $T_K$. The tunneling
contribution to transmission depends very weakly on the bridge's
temperature \cite{Segal}.
This naive approximation is adjusted by maintaining $\Delta T
\rightarrow 0$. Note that a full self consistent formalism (For
example the Keldish- Kadanoff formalism \cite{NEGF}) should
naturally yield the temperature distribution on the junction
without such assumptions.


We define the thermoelectric voltage $\phi|_{I=0}$ as the voltage
necessary for neutralizing the temperature induced current. The
thermopower is the ratio of the potential energy $e\phi$ to the
temperature difference $\Delta T$ under the condition that the
current vanishes
\beq S=-\lim_{\Delta T\rightarrow 0}\frac{e\phi}{k_B\Delta
T}\bigg |_{I=0}. \eeq
We discuss next the limiting behavior of the thermopower ratio for
the different transport mechanisms. The net current is calculated
 by considering separately the elastic (tunneling)
and inelastic (hopping) components,
Eqs.~(\ref{eq:Trans})-(\ref{eq:Thopp}). The tunneling current
reduces to the standard Landauer formula \cite{Landauer}
\beq I_{tunn}=\frac{e}{\pi \hbar} \int d\epsilon_0  A(\epsilon_0)
\left[ f_L(\epsilon_0)- f_R(\epsilon_0)  \right].
\label{eq:Land}
 \eeq
Assuming the transmission is a smooth function of the energy in
comparison to $k_B T_a$, it can be expanded around $\epsilon_F$
%
\bea
 A(\epsilon)&=&A(\epsilon_F)+ \frac{\partial A}{\partial
\epsilon}\bigg \rvert_{\epsilon_F}(\epsilon-\epsilon_F)+
\frac{1}{2!}\frac{\partial^2A}{\partial\epsilon^2}\bigg|_{\epsilon_F}(\epsilon-\epsilon_F)^2
\nonumber\\
&+&\frac{1}{3!}\frac{\partial^3A}{\partial\epsilon^3}\bigg|_{\epsilon_F}(\epsilon-\epsilon_F)^3+
... \label{eq:Aexp}
 \eea
In the linear response regime of small $\Delta T$ and $\phi$ the
Fermi functions in Eq.~(\ref{eq:Land}) are further expanded
linearly
\beq f_L(\epsilon)-f_R(\epsilon)=-\frac{\partial
f(\epsilon,\epsilon_F,T_a)}{\partial \epsilon}\left [
e\phi+\frac{\Delta T}{T_a}(\epsilon-\epsilon_F) \right].
\label{eq:fexp}\eeq
Here the derivative of the Fermi function is calculated at the
average values $T_a$ and $\epsilon_F$. When we consider only the
first two terms in Eq.~(\ref{eq:Aexp}) in conjunction with
Eq.~(\ref{eq:fexp}), we obtain the standard lowest order expression
for electron current due to both electric bias and temperature
gradient \cite{Datta,Zheng,Proetto}
\beq I_{tunn}=-\frac{e^2}{\pi \hbar}A(\epsilon_F) \phi+
\frac{e}{\pi \hbar}\frac{\pi^2k_B^2T_a}{3} \frac{\partial
A}{\partial\epsilon} \bigg|_{\epsilon=\epsilon_F} \Delta T.
\label{eq:Stand} \eeq
The current through the device is zero when the potential
difference is set to
\beq \phi|_{I=0}=
\frac{\pi^2k_B^2T_a}{3e}\frac{\partial(\ln(A))}{\partial
\epsilon}\bigg|_{\epsilon=\epsilon_F} \Delta T. \label{eq:St0}
\eeq
When higher order terms in $A$ are necessary, we utilize the
Sommerfeld expansion \cite{Ashcroft} and get a power law series in
$T_a$ for the thermoelectric potential

\beq \phi|_{I=0}\sim \frac{k_B \Delta T}{e A(\epsilon_F)}\left(
\sum_{n=1,3,5...}\frac{\partial^nA}{\partial
\epsilon^n}\bigg|_{\epsilon_F}(k_BT_a)^n \right).
\label{eq:Stgenr}
 \eeq
We can further consider an explicit expression for $A$.
 If the bridge energies lies well above
$\mu_L$ and $\mu_R$, a perturbative treatment leads to the
superexchange result  \cite{McConnel}
\beq A(\epsilon_0)\sim \frac{V^{2N}}{(E_B-\epsilon_0)^{2N}} \Gamma.
\label{eq:At} \eeq
Here $V$ is the coupling matrix element in the bridge and
$\Gamma$=$\Gamma_K$ is the relaxation rate to the $K$ metal. We
substitute this relation into Eq.~(\ref{eq:Stgenr}) and get the
tunneling contribution to the thermopower
\beq
  -\frac{e\phi|_{I=0}}{k_B\Delta T}
  \sim \sum_{n=1,3,5...}(2 N)^n \frac{  (k_BT_a)^n} {(E_B-\epsilon_F)^n}
  \equiv S_T.
  \label{eq:St} \eeq
%
This expression, though approximate, provides us with the
important features of $S_T$: the {\it inverse } dependence with
energy gap, and its enhancement with size and temperature.

When the gap $(E_B-\epsilon_F)$ is large and the bridge size $N$ is long 
 such that $A$ is practically zero away from the bridge energies,
  i.e. $A(\epsilon)\propto \delta(\epsilon-E_B)$,
the tunneling contribution from $\epsilon_F$ approaches zero.
Electron transmission occurs then mainly via electrons in the
tails of the Fermi distributions in the leads at energies around
$E_B$. In this ballistic regime the current, Eq.~(\ref{eq:Land}),
is zero when $f_L(E_B)=f_R(E_B)$ or
\beq
\beta_L(E_B-\mu_L)=\beta_R(E_B-\mu_R).
\eeq
Using $T_L=T_a+\Delta T/2$, $T_R=T_a-\Delta T/2$,
$\mu_L=\epsilon_F+e\phi/2$ and $\mu_R=\epsilon_F-e\phi/2$, it
reduces to
\beq  -\frac{e\phi|_{I=0}}{k_B\Delta T}=
\frac{E_B-\epsilon_F}{k_BT_a} \equiv S_B. \label{eq:Sb} \eeq
This is the thermopower ratio in the ballistic regime: Electrons
physically populate the molecule, but inelastic effects on the
bridge are neglected. The thermopower in this case scales like
$T_a^{-1}$, in accordance with previous studies \cite{Oreg}.
%

Next we estimate the thermopower when thermal interactions govern
the transport across the bridge. We substitute
Eq.~(\ref{eq:Thopp}) into Eq.~(\ref{eq:curr}) using
$\beta=\beta_K$ for the $\mathcal T_{K\rightarrow K'}$
calculation, and get the net thermal (hopping) current
\bea I_{hopp}&=&\frac{e}{\pi \hbar} \int d\epsilon_f \int
d\epsilon_0
B(\epsilon_0,\epsilon_f)(1-f_L(\epsilon_0))(1-f_R(\epsilon_f))
\nonumber\\
&\times&\left[
 e^{-\beta_L(E_B-\mu_L)}-
 e^{-\beta_R(E_B-\mu_R)}\right].
\eea
It is zero when the term in the square parentheses vanishes, i.e.
\beq \beta_L(E_B-\mu_L)=\beta_R(E_B-\mu_R), \eeq
which leads to
\beq  -\frac{e\phi|_{I=0}}{k_B\Delta T}=
\frac{E_B-\epsilon_F}{k_BT_a} \equiv S_H. \label{eq:Sh} \eeq
This result is equivalent to  Eq. (\ref{eq:Sb}). There we
considered resonant-coherent-transmission through a long chain,
when tunneling from the Fermi energy is negligible. Therefore, we
cannot distinguish  in the thermopower between transport due to
thermal activation from the lead to the molecule and band motion
of electrons at the tails of the Fermi function. We refer to both
as thermal mechanisms.

We compare next the tunneling thermopower term $S_T$, Eq.
(\ref{eq:St}), to the thermalized behavior $S_H$,
(Eqs.~(\ref{eq:Sb}), (\ref{eq:Sh})). The following observations
can be made: (i) $S_T \sim (T_a / \Delta E)^n$, while $S_H\sim
\Delta E/T_a$. Here $\Delta E\equiv E_B-\epsilon_F$. (ii) The
tunneling term depends on the length of the molecule $N$. (iii) In
both cases $S$ does not depend on the molecule-metal interaction
strength, given by the parameter $\Gamma$. (iv) Measurement of $S$
vs. $T_a$ should yield the effective energy gap for transmission,
and  also hint on the transport mechanism. We expect that when
increasing the temperature, the thermopower should first
increase (tunneling behavior governs at low temperatures), then
decay in a $T_a^{-1}$ fashion when thermal activation dominates
electron transfer.

\section{ Results}

We investigate the thermopower behavior in our model system, Eqs.
(\ref{eq:Ham})-(\ref{eq:mu}), and show how the approximate
expressions, Eqs. (\ref{eq:St}) and (\ref{eq:Sh}), agree well with
the numerical results. We also extract important energetic and
dynamic information from the thermoelectric potential.
The following set of parameters is used:  $N \sim$ 1-5 units,
molecular energies  $\tilde E \sim$ 100-500 meV ($\epsilon_F$ is
taken as zero). We focus on the limit $\tilde E> V$  using $V$=25
meV. The metal-molecule relaxation rate is taken equal at the $L$
and $R$ sides, $\Gamma_K=15$ meV. The system bath interaction is
given by the dephasing parameter, $\gamma\sim 0.5-10$ meV.



\begin{figure}[htbp]
\vspace{0mm} \hspace{0mm}
 {\hbox{\epsfxsize=70mm \epsffile{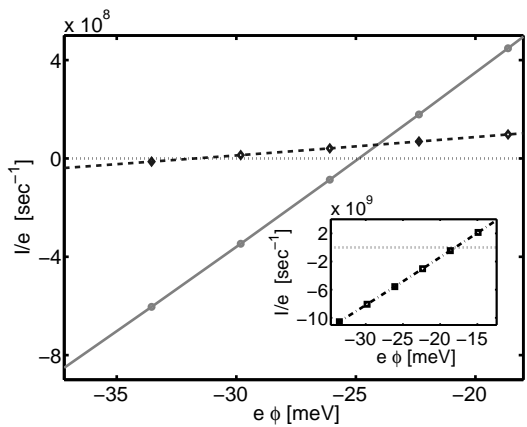}}}
 \caption{ I-V characteristics of the $N$=4 junction.
 The system parameters are $\gamma$=5 meV,
 $T_L$=335 K, $T_R$=300 K.
 $\tilde E=315$ meV (dashed),  $\tilde E=250$
 meV (full).
 The dotted line shows for reference the $I$=0 function.
  Inset: $\tilde E=185$ meV.}
 \label{Fig2}
\end{figure}

\begin{figure}[htbp]
\vspace{0mm} \hspace{0mm}
 {\hbox{\epsfxsize=70mm \epsffile{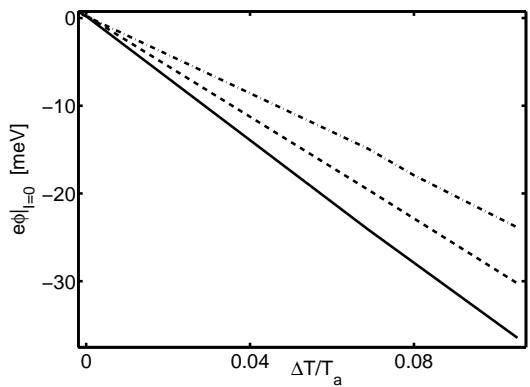}}}
 \caption{ The thermoelectric voltage plotted against $\Delta T/ T_a$ for different
 metal-molecule energy gaps.
   $\tilde E $=375 meV (full),
   $\tilde E $=315 meV (dashed), and
 $\tilde E $=250 meV   (dashed-dotted).
  $N$=4, $\gamma$=5 meV, $\Delta T$= 30 K.
}
 \label{Fig3}
\end{figure}

\begin{figure}[htbp]
\vspace{0mm} \hspace{0mm}
 {\hbox{\epsfxsize=70mm \epsffile{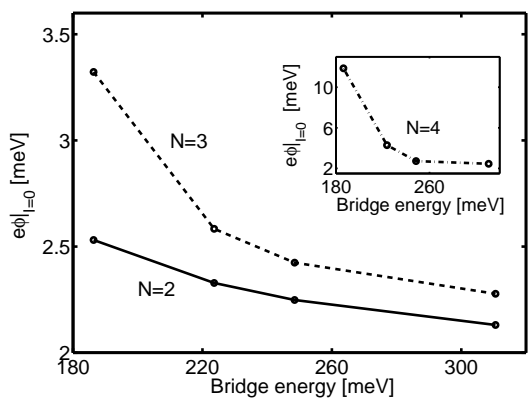}}}
 \caption{ Dependence of the thermoelectric voltage on the bridge energy ($\tilde E$)  in the
 superexcnahge regime.
$\gamma$=0.5 meV, $T_a$=100 K, $\Delta T$= -20 K.
 $N$=2 (full),  $N$=3 (dashed).
 Inset: $N$=4.
}
 \label{Fig4}
\end{figure}

Fig. \ref{Fig2} shows the I-V characteristic for a chain of size $N$=4 for
different energy gaps $\tilde E$ using $T_L$=335 K and $T_R$=300
K. Within this set of parameters the current is dominated by
thermal effects, see Fig.~\ref{Fig0}. Note that the I-V
characteristic is expected to be linear for $e\phi<k_BT_K$ for
both coherent and incoherent modes of transfer \cite{Segalh}. This
property enables easy and accurate evaluation of the $\phi|_{I=0}$
value by a linear fitting of a few I-V data points.

We find that the results perfectly agree with Eq.~(\ref{eq:Sh}).
For $\tilde E$=315 meV (dashed) the current vanishes at
$e\phi|_{I=0}$=-32 meV. When utilizing Eq.~(\ref{eq:Sh}) this
number provides an effective energy gap of $E_B$=290 meV. This
value is in agreement with the lowest diagonalized bridge energy
of $\sim \tilde E-V$=315-25=290 meV. The same behavior is obtained
for $\tilde E$=250 meV (full) where we get $e\phi|_{I=0}$=-25 meV,
leading to $E_B$=227 meV. For $\tilde E$=187 meV (inset) the
zero-current voltage is $e\phi|_{I=0}$=-18 meV and the resulting
gap is $E_B$=163 meV.
We note that the effective energy gap $E_B$ calculated (or
measured) based on the thermoelectric effect is the {\it real} gap in
the system, taking into account naturally and consistently the
metal-molecule interaction, the thermal effects and the applied
bias.


 Fig.~\ref{Fig3} demonstrates the temperature dependence  of the
thermoelectric voltage within the same range of parameters. It
provides another evidence that in the present case transport is
dominated by thermal activation as $\phi_{I=0}\propto \Delta
T/T_a$, in agreement with Eq.~(\ref{eq:Sh}). The slopes for the
$\tilde E=375, 315, 250$ meV cases are 316, 267,  217 meV
 respectively, producing the effective gaps.
Note that for different thermoelectric voltages the bridge
energies are slightly varied according to Eq.~(\ref{eq:mu}). For
accurate results measurements should be done at  $\Delta
T\rightarrow 0$.

We turn now to the low temperature regime where tunneling is
expected to dominate charge transfer. Fig.~\ref{Fig4} presents the
potential $\phi|_{I=0}$ for various chain lengths as a function of
the energy gap $\tilde E$ using $T_a$=100 K and $\gamma$=0.5 meV.
Indeed we find that the thermoelectric voltage {\it decreases}
with increasing gap in contrast to its behavior (in absolute
values) in figures \ref{Fig2} and \ref{Fig3}. In addition, 
the thermoelectric voltage is larger for longer bridges. These
observations are consistent with Eq.~(\ref{eq:St}).



\begin{figure}[htbp]
\vspace{0mm} \hspace{0mm}
 {\hbox{\epsfxsize=70mm \epsffile{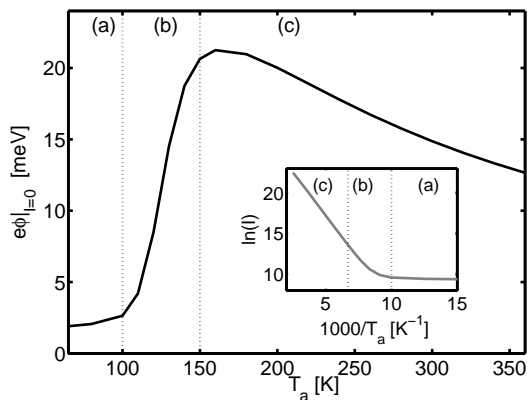}}}
 \caption{ Temperature dependence of the thermoelectric voltage.
  $\tilde E $=250 meV,
  $N=4$, $\gamma$=1 meV, $\Delta T$= -20 K.
 Inset: Arrhenius plot of current vs. inverse temperature for $e\phi=6$ meV.
}
 \label{Fig5}
\end{figure}

Next we show that  the temperature induced turnover between
coherent motion to incoherent transmission is reflected in the
thermoelectric potential. Fig.~\ref{Fig5} exemplifies this
behavior. We observe mainly three regimes in the main curve: (a)
The potential is almost zero. (b) $e\phi|_{I=0}$ increases
strongly with $T_a$. (c) Above the threshold of $T_a \sim$ 150 K
the potential saturates, then goes down like $\propto \Delta
T/T_a$. We can explain these results as follows: (a) At very low
temperatures the thermoelectric potential is close to zero as
tunneling transmission through the bridge is very small. Ballistic
motion is not significant because of the low population of
electrons at the metals around $E_B$. (b) When the temperature is
increased electrons populate energies in the metals above the
Fermi energy and tunneling becomes more probable. Band motion and
thermalized hopping also begin to contribute. (c) In this regime
transport occurs through physical population of the bridge, and
the thermopower behaves in accordance with Eq. (\ref{eq:Sh}).
 To conclude- while in region (a) transport is dominated by coherent
interactions, in region (c) it is induced by thermal
effects. The intermediate area (b) presents a regime where
coherent effects and thermal interactions mix.
The inset displays the Arrhenius plot of Ln current vs. inverse
temperature for a representative applied potential, $e\phi=6$ meV
(The results do not depend on the applied voltage for $e\phi$ up to
$\sim$ 30 meV). The different regimes are marked according to the
main plot. A clear transition at $T_a\sim $ 100 K is observed: The
current becomes temperature independent for lower temperatures
since tunneling dominates. The behavior at region (c) is also
clear: ${\rm Ln}(I) \propto 1/T_a$. In contrast, the behavior at
the central area 
is obscure, and no clear transition is observed at
$T_a\sim 150$ K. Thus, thermopower measurements may complement
standard I-V studies, yielding valuable information about the
junction energetics and charge transfer mechanisms.


\begin{figure}[htbp]
\vspace{0mm} \hspace{0mm}
 {\hbox{\epsfxsize=70mm \epsffile{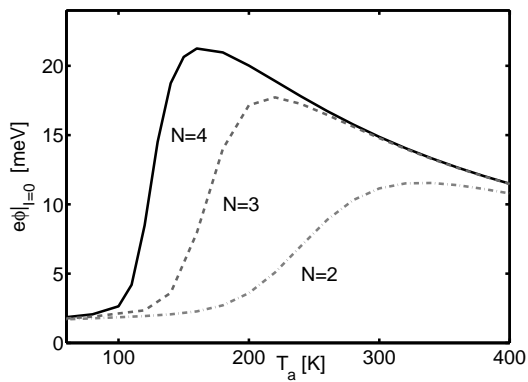}}}
 \caption{Temperature dependence of the thermoelectric voltage for different
  bridge lengths.  $N$=4 (full), $N$=3 (dashed), $N$=2 (dashed-dotted).
  $\tilde E $=250 meV,
   $\gamma$=1 meV, $\Delta T$= -20 K.
}
 \label{Fig6}
\end{figure}

Finally, Fig.~\ref{Fig6} displays the thermoelectric potential vs.
temperature for molecules of different sizes:  $N$=4 (full),
 $N$=3 (dashed), and  $N$=2 (dashed-dotted). We find that for short
chains the transition points between the three regimes are shifted
to higher temperatures as coherent effects persist. For the
$N$=2 chain  transport is still largely controlled by tunneling at
room temperature.
 Another important feature of the thermoelectric potential is its independence on $N$ at high
temperatures, while it is strongly size dependent at low $T_a$.
This observation is  consistent with
expressions (\ref{eq:St}) and (\ref{eq:Sh}).


\section {Summary}

Using a simple model for electron transfer through a molecular
junction, taking into account both coherent effects and thermal
interactions in the molecule, we have calculated the thermopower of the device
assuming weak molecule-leads and weak molecule-thermal baths interactions. 
We have shown that the
thermopower can provide information on charge transfer mechanisms,
and that it can reveal the location of the Fermi energy relative
to the molecular levels. Thermopower measurements can thus
complement standard I-V experiments, with the advantage of their
insensitivity to the metal-molecule contact.
For example, the thermopower can be used  for comparing of
transport experiments done on the same molecules but with
different contact interactions and measurement methods
\cite{Salomon}.

\begin{acknowledgments}
The author would like to thank A. Nitzan and 
M. Shapiro for useful comments. 
This research was supported by a grant from the Feinberg
graduate school of the Weizmann Institute of Science.
\end{acknowledgments}



\begin{thebibliography}{32}



\bibitem{MoleR}
A. Nitzan, M. A. Ratner, Science {\bf 300}, 1384 (2003).

\bibitem{IVexp1}
S. J. Tans {\it et al.},
Nature {\bf 386}, 474 (1997).

\bibitem{IVexp2}
R. H. M. Smit {\it et al.}, Nature {\bf 419}, 906 (2002).

\bibitem{Ratner}
W. B. Davis, W. A. Svec, M. A. Ratner, M. R. Wasielewski,
Nature {\bf 396}, 60 (1998).

\bibitem{DNA}
B. Giese,  J. Amaudrut,  A.-K. K\"ohler, M. Spormann, S. Wessely,
Nature {\bf 412}, 318 (2001).

\bibitem{Selzer}
 Y. Selzer, M. A. Cabassi, T. S. Mayer, D. L. Allara, J. Am. Chem. Soc. {\bf 126}, 4052 (2004); Nanotechnology {\bf 15}, S483 (2004).

\bibitem{Emberly}
E. G. Emberly, G. Kirczenow, Phys. Rev. B {\bf 61}, 5740 (2000).

\bibitem{Ness}
H. Ness, S. A. Shevlin, A. J. Fisher, Phys. Rev. B {\bf 63},
125422 (2001);
H. Ness, A. J. Fisher, Phys. Rev. Lett. {\bf 83}, 452 (1999).

\bibitem{Segal}
D. Segal, A. Nitzan, W. B. Davis, M. R. Wasielewski, M. A. Ratner,
J. Phys. Chem. B {\bf 104}, 3817 (2000); D. Segal, A. Nitzan, M.
A. Ratner, W. B. Davis, J. Phys. Chem. {\bf 104}, 2790 (2000); D.
Segal, A. Nitzan, Chem. Phys. {\bf 268}, 315 (2001); {\bf 281},
235 (2002).

\bibitem{Todorov}
M. J. Montgomery, J. Hoekstra, T. N. Todorov, A. P. Sutton, J.
Phys.:
 Condens. Matter {\bf 15}, 731 (2003).

\bibitem{Petrov}
E. G. Petrov, V. May, J. Phys. Chem. A  {\bf 105}, 10176 (2001).
E. G. Petrov, Y. R. Zelinskyy, V. May, J. Phys. Chem. B {\bf 106},
3092 (2002); E. G. Petrov, V. May, P. H\"anggi, Chem. Phys. {\bf
296}, 251 (2004).


\bibitem{Datta}
M. Paulsson, S. Datta, Phys. Rev. B {\bf 67}, 241403 (2003).

\bibitem{Zheng}
X. Zheng, W. Zheng, Y. Wei, Z. Zeng, J. Wang, J. Chem. Phys. {\bf
121}, 8537 (2004).


\bibitem{Oreg}
J. Koch, F. Von Oppen, Y. Oreg, E. Sela, Phys. Rev. B {\bf 70},
195107 (2004).


\bibitem{QD}
C. W. J. Beenakker, A. A. M. Staring, Phys. Rev. B {\bf 46}, 9667
(1992).

\bibitem{SET}
K. A. Matveev, A. V. Andreev, Phys. Rev. B {\bf 66}, 045301
(2002).

\bibitem{CNT1}
J. Hone {\it et al.}, Phys. Rev. Lett. {\bf 80}, 1042 (1998).
\bibitem{CNT2}
J. P. Small, K. M. Perez, P. Kim, Phys. Rev. Lett. {\bf 91},
256801 (2003).

\bibitem{Landauer}
R. Landauer, Philos. Mag. {\bf 21}, 863 (1970); R. Landauer, IBM
J. Res. Dev. {\bf 1}, 223 (1957).


\bibitem{bonca} K. Haule, J. Bon\u ca, Phys. Rev. B {\bf 59}, 13087
(1999).

\bibitem{Nitzan} A. Nitzan, Annu. Rev. Phys. Chem. {\bf 52}, 681
(2001).
%
%
\bibitem{antonyuk}
V. B. Antonyuk, A. G. Mal'shukov, M. Larsson,  K. A. Chao, Phys.
Rev. B {\bf69}, 155308 (2004).
%
\bibitem{MishaJPC}
M. Galperin, A. Nitzan, M. A. Ratner, D. R. Stewart, J. Phys.
Chem. B {\bf 109}, 8519 (2005).

%
\bibitem{domcke}
M. \u Ci\u zek, M. Thoss, W. Domcke, Phys. Rev B {\bf 70}, 125406
(2004).

%
\bibitem{Segalh}
D. Segal, A. Nitzan, J. Chem. Phys. {\bf 117}, 3915 (2002).
%

\bibitem{Segalc}
D. Segal, A. Nitzan, P. H\"anggi, J. Chem. Phys. {\bf 119}, 6840
(2003).

\bibitem{NEGF}
L. P. Kadanoff and G. Baym, Quantum Statistical Mechanics.
(Benjamin-Cummings, Reading, MA, 1962); L. V. Keldysh, Sov. Phys.
JETP {\bf 20}, 1018 (1965).

\bibitem{Proetto}
C. R. Proetto, Phys. Rev. B {\bf 44}, 9096 (1991).

\bibitem{Ashcroft}
N. W. Ashcroft, N. D. Mermin, Solid State Physics (Saunders
College publishing, Philadelphia 1976).

\bibitem{McConnel}
H. M. McConnell, J. Chem. Phys. {\bf 35}, 508 (1961).

\bibitem{Salomon}
A. Salomon {\it et al.}, Advanced Materials {\bf 15}, 1881 (2003).

\end{thebibliography}
\end{document}